\documentclass[journal=nalefd,manuscript=article,layout=twocolumn]{achemso}

\setkeys{acs}{etalmode=truncate,maxauthors=5,usetitle=true,manuscript=letter}

\usepackage[version=3]{mhchem} % Formula subscripts using \ce{}

\newcommand*{\mycommand}[1]{\texttt{\emph{#1}}}

\usepackage{hyperref}
\hypersetup{colorlinks}
\usepackage{color}
\definecolor{darkred}{rgb}{0.5,0,0}
\definecolor{darkgreen}{rgb}{0,0.5,0}
\definecolor{darkblue}{rgb}{0,0,0.5}
\hypersetup{ colorlinks,
linkcolor=blue,
filecolor=darkgreen,
urlcolor=darkred,
citecolor=red }
\usepackage{braket}
\usepackage{lineno}

\author{Neerav Kharche}
 \email{kharcn@rpi.edu}
 \affiliation{Computational Center for Nanotechnology Innovations, Rensselaer Polytechnic Institute, Troy, NY 12180, USA}
 \altaffiliation{Department of Physics, Applied Physics and Astronomy, Rensselaer Polytechnic Institute, Troy, NY 12180, USA} 

  \author{Saroj K. Nayak}
 \email{nayaks@rpi.edu}
 \affiliation{Department of Physics, Applied Physics and Astronomy, Rensselaer Polytechnic Institute, Troy, NY 12180, USA} 

\date{\today}

\title[\texttt{achemso} demonstration]
{Quasiparticle bandgap engineering of graphene and graphone on hexagonal boron nitride substrate}

\begin{document}
%\linenumbers

%%%%%%%%%%%%%%%%%%%%%%%%%%%%%%%%%%%%%%%%%%%%%%%%%%%%%%%%%%%%%%%%%%%%%
%% The manuscript does not need to include \maketitle, which is
%% executed automatically.  The document should begin with an
%% abstract, if appropriate.  If one is given and should not be, the
%% contents will be gobbled.
%%%%%%%%%%%%%%%%%%%%%%%%%%%%%%%%%%%%%%%%%%%%%%%%%%%%%%%%%%%%%%%%%%%%%

\begin{abstract}
Graphene holds great promise for post-silicon electronics, however, it faces two main challenges: opening up a bandgap and finding a suitable substrate material. In principle, graphene on hexagonal boron nitride (hBN) substrate provides potential system to overcome these challenges. Recent theoretical and experimental studies have provided conflicting results: while theoretical studies suggested a possibility of a finite bandgap of graphene on hBN, recent experimental studies find no bandgap. Using the first-principles density functional method and the many-body perturbation theory, we have studied graphene on hBN substrate. A Bernal stacked graphene on hBN has a bandgap on the order of $0.1 \ \rm eV$, which disappears when graphene is misaligned with respect to hBN. The latter is the likely scenario in realistic devices. In contrast, if graphene supported on hBN is hydrogenated, the resulting system (graphone) exhibits bandgaps larger than $2.5 \ \rm eV$. While the bandgap opening in graphene/hBN is due to symmetry breaking and is vulnerable to slight perturbation such as misalignment, the graphone bandgap is due to chemical functionalization and is robust in the presence of misalignment. The bandgap of graphone reduces by about $1 \ \rm eV$ when it is supported on hBN due to the polarization effects at the graphone/hBN interface. The band offsets at graphone/hBN interface indicate that hBN can be used not only as a substrate but also as a dielectric in the field effect devices employing graphone as a channel material. Our study could open up new way of bandgap engineering in graphene based nanostructures. \\\\
%PACS numbers 73.22.-f, 73.22.Pr
%\vfill{\bf \today}
\end{abstract}
%\maketitle


%% ------------------------------ TOC graphic --------------------------------------
%\begin{tocentry}
%\includegraphics{TOC_BN_Graphene_Graphone_2}
%Substrate polarization induced bandgap narrowing in semi-hydrogenated graphene (graphone).
%\end{tocentry}
%% ------------------------------ Figure 1 --------------------------------------

%\section{Introduction}
%\label{sec:intro}
Graphene exhibits remarkable electronic properties compared to the conventional materials such as Si and III-Vs making it an attractive material for next generation electronic devices \cite{Schwierz_NNano_2010}. For practicle applications, devices made from an atomically thin material such as graphene should be supported on a substrate. Typically graphene devices are fabricated on $\rm SiO_2$ substrate, however, carrier mobility in graphene on $\rm SiO_2$ reduces due to charged surface states, surface roughness, and surface optical phonons in $\rm SiO_2$ \cite{Chen_NNano_2008,Ponomarenko_PRL_2009}. Several other oxide-based substrates have been investigated so far, however, none yields significant improvement over $\rm SiO_2$ \cite{Ponomarenko_PRL_2009}. Recently, graphene supported on a hexagonal boron nitride (hBN) substrate was found to exhibit much higher mobility compared to any other substrate \cite{Dean_NNano_2010,Xue_NMat_2011}. High mobility of graphene on hBN is enabled by extremely flat surface of hBN and abscence of dangling bonds at the graphene/hBN interface \cite{Xue_NMat_2011}.

Another important challenge for the use of graphene in devices is the lack of controllable bandgap \cite{Schwierz_NNano_2010}. A bandgap can be opened through quantum confinement by patterning graphene into the so-called graphene nano-ribbons (GNRs) \cite{Han_RPL_2007,Son_PRL_2006}. However, it is difficult to control the bandgap in GNRs due to its sensitivity to the width and edge geometry \cite{Son_PRL_2006}. Alternatively, the bandgap can be opened by chemical functionalization of graphene with a variety of species such as H, F, OH, etc \cite{Li_ACSNano_2011}. The hybridization of the functionalized C atom changes from $sp^2$ to $sp^3$, which opens up a bandgap. The bandgap opened by this mechanism is expected to be more robust in the presence of disorder compared to the bandgap opened by the quantum confinement \cite{Abanin_PRL_2010}. The bandgap opening by H-functionalization/hydrogenation of graphene has been a subject of several recent experimental and theoretical studies, which show that the bandgap of graphene can be tuned by controlling the degree of hydrogenation \cite{Balog_NMat_2010,Haberer_NanoLetters_2010,Elias_Science_2009,Zhou_NanoLetters_2009,Fiori_PRB_2010}.

Here we report the electronic structure of graphene and single-sided hydrogenated graphene (graphone) supported on the hBN substrate calculated using the first-principles density functional method and the many-body perturbation theory in the GW approximation, the state-of-the-art method for accurate predictions of the electronic structure. Theoretical studies have suggested a possibility of inducing a bandgap in graphene when supported on the hBN substrate \cite{Giovannetti_PRB_2007}, however, recent experimental studies find no bandgap in this system \cite{Dean_NNano_2010,Xue_NMat_2011}. Earlier calculations based on the tight-binding model ascribe this discrepancy to the random stacking arrangement of graphene on hBN \cite{Xue_NMat_2011}. The tight-binding model can not be reliably used for quantitative predictions involving novel materials such as graphene on hBN especially because the interlater hopping parameters between graphene and hBN are not known. First-principles calculations are therefore required for accurate prediction of electronic structure of such novel material systems. Our calculations show that slight misalignment of graphene closes the bandgap induced by hBN in Bernal stacked graphene. We consider hydrogenation as an alternative to induce bandgap in graphene supported on hBN. Hydrogenation of a substrate supported graphene results in $50 \%$ hydrogen coverage \cite{Subrahmanyam_PNAS_2011} and the resulting material is called as graphone \cite{Zhou_NanoLetters_2009,Li_ACSNano_2011}. Free standing graphone exhibits a bandgap larger than $2.5 \ \rm eV$. Interestingly, the bandgap of graphone reduces by about $1 \ \rm eV$ due to the substrate induced polarization effects \cite{Neaton_PRL_2006,Thygesen_PRL_2009,Freysoldt_PRL_2009,Li_JChemTheoComp_2009} when it is supported on hBN. However, unlike graphene, the bandgap of graphone is unaffected by the misalignment with respect to hBN. The band offsets at graphone/hBN interface suggest that hBN can be used not only as a substrate but also as a dielectric in the field effect devices employing graphone as a channel material.

%\section{Model geometry and computational methods}

%\section{Methods}
The electronic structure calculations are performed in the framework of density functional theory (DFT) within the local density approximation (LDA) as implemented in the ABINIT code \cite{Gonze_CompPhysComm_2009}. The Trouiller-Martins norm-conserving pseudopotentials \cite{Troullier_PRB_1991} and the Teter-Pade parameterization for the exchange-correlation functional \cite{Goedecker_PRB_1996} are used. To ensure negligible interaction between periodic images, a large value ($10 \ $ \AA) of the vacuum region is used. The Brillouin zone is sampled using Monkhorst-Pack meshes of different size depending on the size of the unit cell: $18\times18\times1$ for Bernal stacked graphene on hBN, $6\times6\times1$ for misaligned graphene on hBN, $18\times18\times1$ for chair-graphone, and $8\times8\times1$ for boat-graphone. For the plane wave expansion of the wavefunction, a $30 \ \rm Ha$ kinetic energy cut-off is used. The quasiparticle corrections to the LDA bandstructure are calculated within the $\rm G_0W_0$ approximation and the screening is calculated using the plasmon-pole model \cite{Hybertsen_PRB_1986}.

The Vienna \textit{ab initio} simulation package (VASP) \cite{kresse1996_VASP1}, which provides well-tested implementation of van der Waals interactions (vdW) \cite{Bucko_JPCA_2010}, is used to compute the equilibrium distance between graphene (or graphone) and hBN. The PAW pseudopotentials \cite{Kresse_PRB_1999_VASP_PAW}, the PBE exchange-correlation functional in the GGA approximation \cite{Perdew_PRL_1996_PBE}, and the DFT-D2 method of Grimme \cite{Grimme_JCC_2006} are used. Same values of energy cutoff, vacuum region, and k-point grid as in the ABINIT calculations are used in VASP calculations. The optimized geometries calculated using VASP and ABINIT without including vdW interactions are found to agree well with each other.

% ------------------------------ Figure 1 --------------------------------------
%\newpage
\begin{figure}[t]
%\begin{center}
\resizebox{3.3in}{!}{
\includegraphics{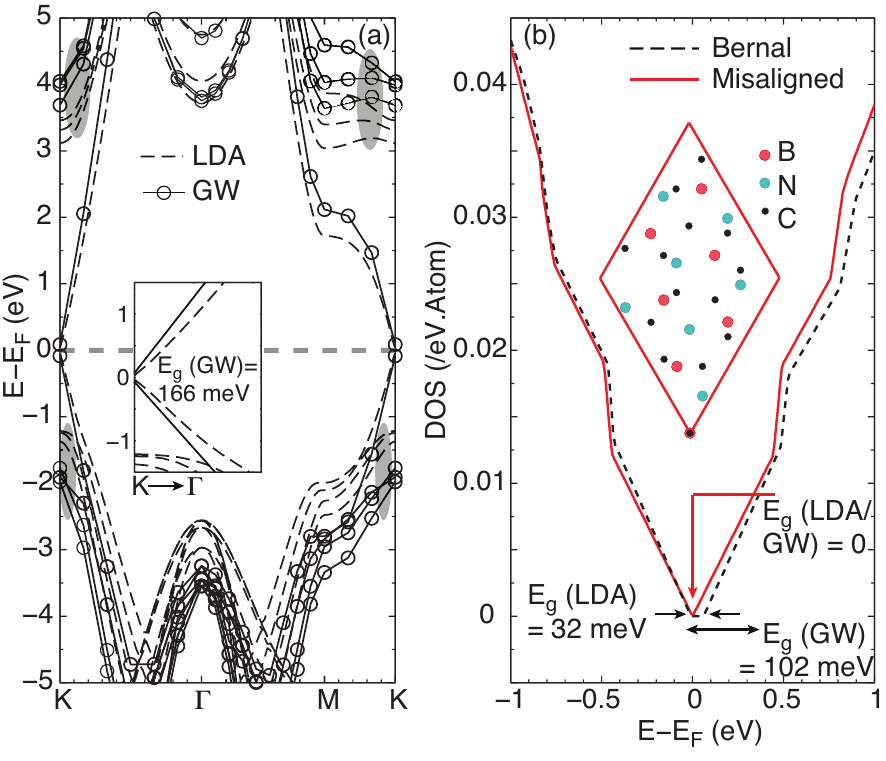}
}
%\end{center}
\caption{(a) The LDA and GW bandstructures of Bernal stacked graphene on hBN substrate. The inset shows a small bandgap opening at the \textit{K} point. (b) Bandgap closing due to misalignment illusrated by the LDA density of states and GW bandgaps of the heterogeneous bilayer of graphene and hBN in perfect Bernal and misaligned stacking arrangements. The inset shows atomistic schematic of a commensurate unit cell of the misaligned bilayer.}
\label{fig:bands_and_Bernal_misaligned_DOS}
\end{figure}
% ------------------------------ Figure 1 --------------------------------------

%\section{Results and discussion}

Figure~\ref{fig:bands_and_Bernal_misaligned_DOS}(a) shows LDA and GW bandstructures of Bernal stacked graphene on hBN substrate such that half of the C atoms in graphene are positioned exactly above the B atoms. This stacking arrangement has been found to be lowest energy configuration \cite{Giovannetti_PRB_2007}. The equlibrium distance between graphene and hBN is $3.14 \ $ \AA. Three monolayers of the semi-infinite hBN substrate are included in the simulation domain to ensure that the GW bandgap is converged. The bands contributed by hBN are identified separately. The weak interlayer interaction between graphene and hBN allows graphene to retain its linear bandstructure near the $K$ point. Underlying hBN substrate induces sublattice asymmetry on the graphene lattice opening up a small bandgap at the $K$ point as shown in the inset of Figure~\ref{fig:bands_and_Bernal_misaligned_DOS}(a). The LDA bandgap is smaller by about $0.1 \ \rm eV$ compared to the GW corrected bandgap.

Contrary to the above conclusion and earlier theoretical studies, recent transport measurements on graphene supported on hBN show no evidence of bandgap \cite{Dean_NNano_2010,Xue_NMat_2011}. The perfect Bernal stacking of graphene on hBN is difficult to achieve in the experiments and random orientation is more probable. To mimic the random orientation, we simulate larger supercells where the stacking between the graphene layer and the underlying hBN substrate deviates from the ideal Bernal stacking. To reduce the computational requirements only one layer of hBN is included. In the Bernal stacking, the graphene layer is rotated by an angle $\theta = 30\,^{\circ}$ with respect to hBN. We consider three rotation angles $21.8\,^{\circ}$, $32.2\,^{\circ}$, and $13.2\,^{\circ}$. The commensurate supercells for these rotations contain 28 (14 C, 7 B, 7 N), 52 (26 C, 13 B, 13 N), and 76 (38 C, 19 B, 19 N) atoms respectively. The supercell for $21.8\,^{\circ}$ rotation is depicted in the inset of Figure~\ref{fig:bands_and_Bernal_misaligned_DOS}(b). The remaining two supercells and the Brillouin zones of all three supercells are shown in the supplementary material.  The commensuration conditions derived in \cite{PRL_Shallcross_2008} are used to generate the supercells. In these supercells, the sublattice asymmetry induced by the hBN substrate on the graphene layer is significantly reduced compared to the Bernal stacking.

The LDA density of states of the misaligned graphene ($\theta = 21.8\,^{\circ}$) is compared with that of the Bernal stacked graphene in Figure~\ref{fig:bands_and_Bernal_misaligned_DOS}(b). The Bernal stacked graphene has a finite bandgap, which closes due to the slight misalignment. Since LDA is known to underestimate the bandgap, we have calculated the GW corrections to the bandgap of misaligned graphene. The GW corrected bandgap of the Bernal stacked graphene increases from $E_g^{LDA} = 68 \ \rm meV$ to $E_g^{GW} = 145 \ \rm meV$ while the GW corrected bandgap of the misaligned graphene remains $0$. The GW bandgaps for rotations $32.2\,^{\circ}$ and $13.2\,^{\circ}$ are also $0$. The Moir\'{e} patterns larger than those shown in Figure~\ref{fig:bands_and_Bernal_misaligned_DOS}(b) were recently observed in graphene supported on hBN \cite{Xue_NMat_2011}. These graphene samples indeed showed $0$ bandgap similar to our calculations.

% ------------------------------ Figure 2 --------------------------------------
\begin{figure}[t]
%\begin{center}
\resizebox{3.3in}{!}{

\includegraphics{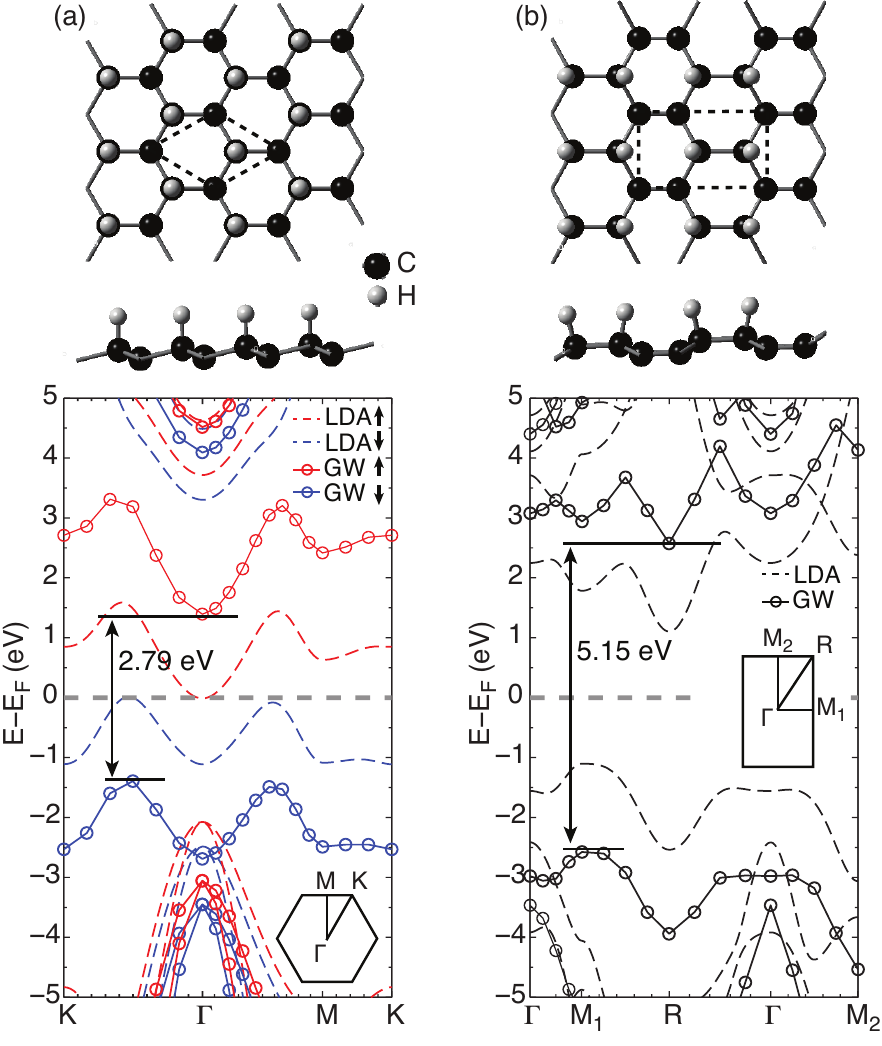}

}
%\end{center}
\caption{Atomistic schematics and bandstructures of graphone in (a) chair and (b) boat conformations. Unit cells are depicted by the dotted lines in the top views of atomistic schematics while Brillouin zones are shown in the insets of bandstructure plots.}
\label{fig:unpaired_paired_H2_graphene_bands}
\end{figure}
% ------------------------------ Figure 2 --------------------------------------

The above calculations indicate that due to misalignment hBN substrate can not reliably induce a bandgap in graphene and alternative approaches are required to open up a sizable bandgap. Recent experimental measurements and theoretical calculations show that the bandgaps on the order of $1 \ \rm eV$ can be opened up in hydrogenated graphene \cite{Balog_NMat_2010,Haberer_NanoLetters_2010,Elias_Science_2009,Zhou_NanoLetters_2009}. Furthermore, the bandgap can be tuned by controlling the degree of hydrogenation. Now, the question arises, what is the effect of hBN substrate on the electronic structure of hydrogenated-graphene? To address this issue, we consider single-sided semi-hydrogenated graphene, which is referred to as graphone \cite{Zhou_NanoLetters_2009}. Recent experimental and theoretical study has shown that hydrogenation of a substrate supported graphene results in $50 \%$ hydrogen coverage \cite{Subrahmanyam_PNAS_2011}, which we have used for the present study. Graphone has two distinct configurations chair (Figure~\ref{fig:unpaired_paired_H2_graphene_bands}(a)) and boat (Figure~\ref{fig:unpaired_paired_H2_graphene_bands}(b)), depending on the relative placement of hydrogenated C atoms. Both chair and boat configurations are energetically stable \cite{Subrahmanyam_PNAS_2011,Li_ACSNano_2011,Zhou_NanoLetters_2009} and they can be synthesized by different hydrogenation processes. A direct single-sided hydrogenation of graphene is likely to result in boat-graphone, which is more stable compared to chair-graphone \cite{Subrahmanyam_PNAS_2011,Li_ACSNano_2011}. On the other hand, chair-graphone can be synthesized by selectively desorbing hydrogen from one side of graphane \cite{Sofo_PRB_2007,Zhou_APL_2009}. The hybridization of C atom changes from $sp^2$ to $sp^3$ upon hydrogenation and the planar structure of graphene becomes non-planar as depicted in Figure~\ref{fig:unpaired_paired_H2_graphene_bands}. 

Graphene is weakely bonded to hBN by van der Waals interaction and the frontier orbitals of graphene, which take part in bonding with hydrogen are virtually unaffected by hBN. Therefore the stable configurations of graphone are likely to be unaffected by hBN. Indeed, the optimized atomic structures of graphone on hBN are virtually identical to earlier studies \cite{Subrahmanyam_PNAS_2011,Li_ACSNano_2011,Zhou_NanoLetters_2009}, which did not include any substrate. The situation is different with the substrates, which have stronger interaction with graphene. For example, graphene on Ir(111) substrate exhibits specific hydrogenation patterns based on the local orientation of graphene with respect to Ir(111) surface \cite{Balog_NMat_2010}. Hydrogen atoms may desorb or diffuse through graphone into the hBN substrate at elevated temperatures thereby affecting the local electronic struture of graphone. However, such phenomena require in depth analysis using methods such as the molecular dynamics and are out of the scope of the present work.

Chair and boat-graphone exhibit very different electronic and magnetic properties as evident from their bandstructures (Figure~\ref{fig:unpaired_paired_H2_graphene_bands}). The ground state of chair-graphone is ferromagnetic while boat-graphone has a nonmagnetic ground state. The bandstructure of chair-graphone is highly spin polarized with a spin splitting of $2.79 \ \rm eV$, which is also its bandgap. This feature makes chair-graphone an attractive material for spintronics \cite{Li_ACSNano_2011}. Boat-graphone has a spin degenerate bandstructure and behaves as an insulator with a large bandgap of $5.15 \ \rm eV$. Hydrogen atoms may migrate over the graphone lattice resulting in the combination of chair- and boat-configurations. In such structures magnetic ordering of chair-graphone may not be preserved. The bandgap is, however, expected to persist \cite{Abanin_PRL_2010}.

% ------------------------------ Figure 3 --------------------------------------
%\newpage
\begin{figure}[t]
%\begin{center}
\resizebox{3.3in}{!}{
\includegraphics{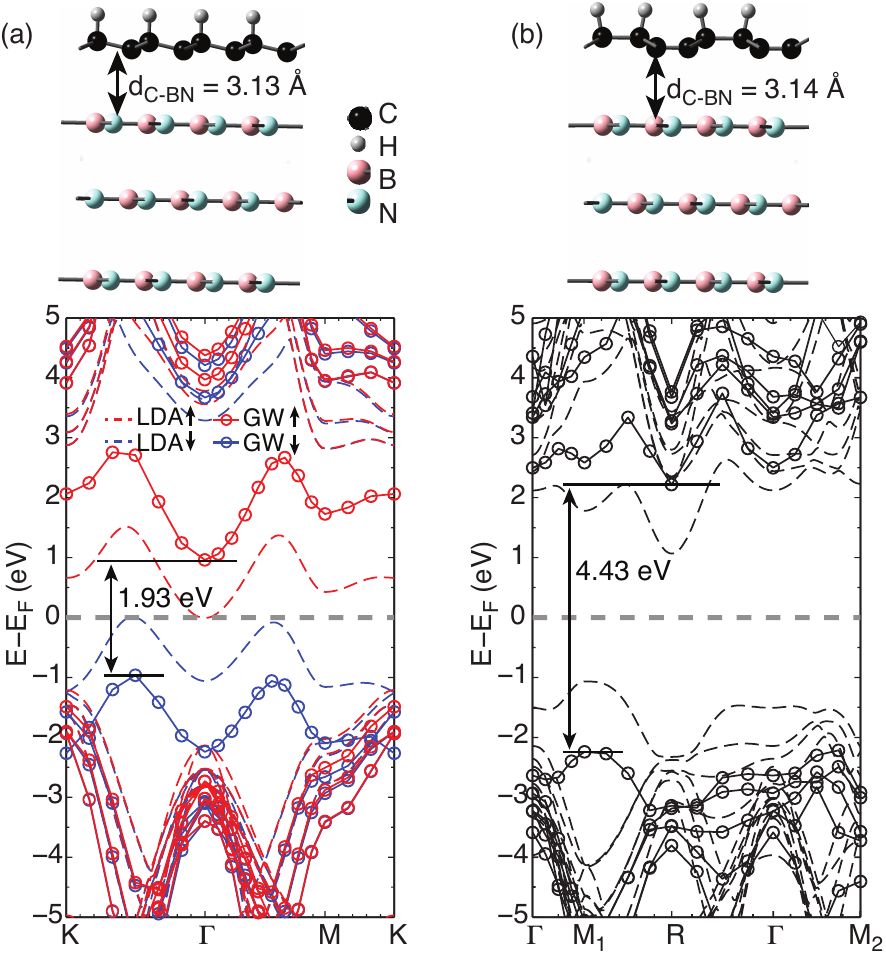}
}
%\end{center}
\caption{Atomistic schematics and bandstructures of (a) chair-graphone and (b) boat-graphone supported on hBN substrate. The substrate polarization induced renormalization of energy levels reduces the bandgap of substrate-supported graphone compared to the free-standing graphone. }
\label{fig:unpaired_paired_H2_graphene_with_BN3_bands}
\end{figure}
% ------------------------------ Figure 3 --------------------------------------

The electronic structure of an atomically thin material such as graphone, when it is used in devices, is expected to be highly dependent on the surrounding materials. The atomistic schematics of graphone supported on hBN substrate and calculated bandstructures of graphone-hBN supercells are shown in Figure~\ref{fig:unpaired_paired_H2_graphene_with_BN3_bands}. The graphone LDA bands in the vicinity of the valence-band maximum and the conduction-band minimum remain unaffected by the presence of hBN substrate. The shape of GW-corrected bands remains more or less unaffected, however, the bandgap of chair and boat-graphone when supported on hBN substrate reduces by $0.86 \ \rm eV$ and $0.72 \ \rm eV$ respectively compared to the bandgap of their free-standing counterparts.

The reduction in GW bandgap is attributed to the polarization effects at the graphone/hBN interface. Similar reductions have been found in experiments and GW calculations for several different molecules adsorbed on metals, semiconductors, and insulators \cite{Neaton_PRL_2006,Thygesen_PRL_2009,Freysoldt_PRL_2009,Li_JChemTheoComp_2009}. Careful inspection of Figures~\ref{fig:unpaired_paired_H2_graphene_bands} and ~\ref{fig:unpaired_paired_H2_graphene_with_BN3_bands} indicates that the GW corrections to the LDA bands are smaller in graphone on hBN compared to those in free standing graphone. This is because of the fact that the polarization of hBN substrate reduces the screened Coulomb potential, $W$, which in turn reduces the GW bandgap. Similar bandgap reduction due to the polarization of hBN substrate is expected in hydrogenated graphene with different hydrogen coverage. The bandgaps of GNRs are also reduced when they are supported on hBN substrate \cite{Jiang_to_be_published}. Non-local polarization effects can not be modeled in the LDA. The GW approach includes non-local polarization effects through the screened Coulomb interaction, however, it tends to be computationally extremely demanding.

The bandgap reduction due to hBN substrate polarization can be estimated by using computationally much less demanding image-charge model \cite{Li_JChemTheoComp_2009}. In this model, the quasiparticle energy of a substrate-supported layer is given by $E_{j;supported}^{QP} = E_{j;free}^{GW} + \Delta P_j$ where $E_{j;free}^{GW}$ is the quasiparticle GW energy of state $\Ket{j}$ of free standing layer, graphone in this case, and $\Delta P_j$ is the correction due to substrate polarization. The quasiparticle bandgaps of chair and boat-graphone estimated using the image-charge model are $2.03 \ \rm eV$ and $4.36 \ \rm eV$ respectively, which compare well with the bandgaps obtained using computationally extensive GW calculations including hBN substrate (Figure~\ref{fig:unpaired_paired_H2_graphene_with_BN3_bands}). The details of image-charge model calculations and underlying assumptions are discussed in the supplemental material.

\begin{figure}[t]
%\begin{center}
\resizebox{3.3in}{!}{
\includegraphics{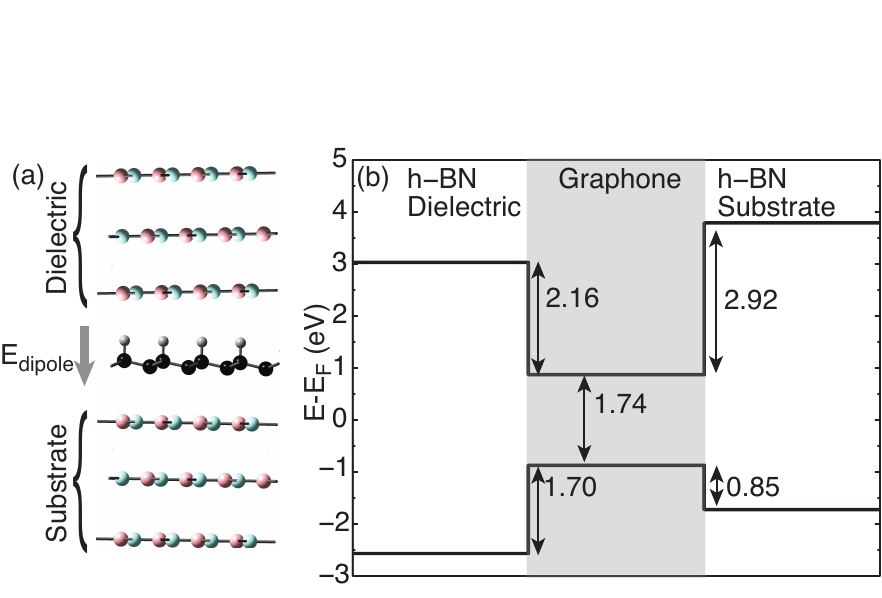}
}
%\end{center}
\caption{Band offsets of chair-graphone embedded in hBN layers. (a) A supercell containing 3 monolayers of hBN on both sides of chair-graphone. The direction of electric field induced by the dipoles in graphone layer is indicated by an arrow. (b) Band offsets at the graphone/hBN interfaces.}
\label{fig:unpaired_paired_H2_graphene_BN_band_offsets}
\end{figure}

In addition to the bandgaps, the band offsets are of crucial importance for the successful use of a heterostructure in the electronic devices. As an illustration, we analyse the conduction and valence band offsets, denoted by $\Delta E_c$ and $\Delta E_v$, respectively, at the chair-graphone/hBN interface when hBN is placed on both the hydrogenated and non-hydrogenated sides of graphone. Figure~\ref{fig:unpaired_paired_H2_graphene_BN_band_offsets}(b) shows a schematic depicting the calculated band offsets in the hBN/chair-graphone/hBN heterostructure. Three monolayers of hBN are included on either sides of graphone. The band offsets do not change when more than 3 monolayers of hBN are included.

% ------------------------------ Figure 4 --------------------------------------
%\newpage
% ------------------------------ Figure 4 --------------------------------------

The bandgap of chair-graphone on hBN reduces from $1.93 \ \rm eV$ (Figure~\ref{fig:unpaired_paired_H2_graphene_with_BN3_bands}) to $1.74 \ \rm eV$ in the presence of hBN on the hydrogenated side of graphone. This is due to the additional reduction of screened Coulomb interaction induced by the polarization of hBN layers on the hydrogenated side. The band offsets at chair-graphone/hBN interfaces are asymmetric such that the bands of hBN on the hydrogenated side of graphone are lower in energy compared to the bands of hBN on the non-hydrogenated side. To investigate the origin of this asymmetry, we carried out Bader charge analysis of a free standing chair-graphone. In chair-graphone, the hydrogenated C-atoms aquire a slight net positive charge while the non-hydrogenated C-atoms aquire a slight net negative charge creating a dipole layer with an electric field depicted by a thick grey arrow in Figure~\ref{fig:unpaired_paired_H2_graphene_BN_band_offsets}(a). This electric field causes asymmetry in the band offsets at graphone/hBN interfaces. Similar modifications of the band offsets have been found in conventional semiconductor homo- and heterojunctions when a dipole layer is inserted at the interface \cite{Pan_APL_1998}.

In field effect devices employing chair-graphone as channel material, hBN can be used as a substrate as well as a dielectric layer separating the gate electrode and the graphone channel. The band diagrams in Figure~\ref{fig:unpaired_paired_H2_graphene_BN_band_offsets}(b) indicate that hBN can be placed on either side of graphone when used as a substrate. When hBN is used as a dielectric layer, however, it should be placed on the hydrogenated side and not on the non-hydrogenated side of graphone. This is because of the fact that when hBN is placed on the non-hydrogenated side of graphone, $\Delta E_v$ at the interface is too low ($< 1 \ \rm eV$) to prevent the thermionic emission of holes into the hBN dielectric layer, which consequently results in gate leakage current degrading the performance of the field-effect device \cite{Robertson_JVCTB_2000}. On the contrary, when hBN is placed on the hydrogenated side of graphone, it has sufficiently high barriers for both electrons ($\Delta E_c$) and holes ($\Delta E_v$) to prevent their thermionic emission into the hBN dielectric layer.

%\section{Conclusion}

In summary, our first-principles calculations suggest that in a Bernal stacked graphene on hBN substrate a bandgap on the order of $0.1 \ \rm eV$ opens up due to the sublattice asymmetry in graphene induced by hBN. This bandgap closes in the presence of slight misorientation from the Bernal stacking. Hydrogenation of graphene provides a promising approach to open up bandgaps larger than $2.5 \ \rm eV$. The polarization effects due to surrounding hBN dielectrics, however, reduce bandgaps of graphone by about $1 \ \rm eV$. Thus, accurate electronic structure calculations of the atomically thin materials such as graphone should always take into account the surrounding materials. The calculated band offsets suggest that in the field effect devices employing chair-graphone as a channel material, hBN can be used as a substrate as well as a dielectric layer separating the graphone channel and the gate electrode. Bandgap engineering of graphene by chemical functionalization is currently an extremely active area of research and this work provides theoretical instruction for analyzing the experimental observations such as the electronic structure modulation by the substrate.

%\begin{acknowledgments}
\acknowledgement
We thank Prof. Timothy Boykin, Prof. Mathieu Luisier, and Prof. Gerhard Klimeck for helpful discussions. This work is supported partly by the Interconnect Focus Center funded by the MARCO program of SRC and State of New York, NSF PetaApps grant number 0749140, and an anonymous gift from Rensselaer. Computing resources of the Computational Center for Nanotechnology Innovations at Rensselaer partly funded by State of New York and of nanoHUB.org funded by the National Science Foundation have been used for this work.
%\end{acknowledgments}

%%\section{Author contributions}
%%Author contributions here.
%%
%%\section{Competing Financial Interests}
%%The authors declare no competing financial interests.

%----------------------------- Bibliography --------------------------------
%\newpage
%\bibliographystyle{plain}
%\bibliography{../BN_Graphene_bibliography_edited}

\begin{mcitethebibliography}{36}
\providecommand*\natexlab[1]{#1}
\providecommand*\mciteSetBstSublistMode[1]{}
\providecommand*\mciteSetBstMaxWidthForm[2]{}
\providecommand*\mciteBstWouldAddEndPuncttrue
  {\def\EndOfBibitem{\unskip.}}
\providecommand*\mciteBstWouldAddEndPunctfalse
  {\let\EndOfBibitem\relax}
\providecommand*\mciteSetBstMidEndSepPunct[3]{}
\providecommand*\mciteSetBstSublistLabelBeginEnd[3]{}
\providecommand*\EndOfBibitem{}
\mciteSetBstSublistMode{f}
\mciteSetBstMaxWidthForm{subitem}{(\alph{mcitesubitemcount})}
\mciteSetBstSublistLabelBeginEnd
  {\mcitemaxwidthsubitemform\space}
  {\relax}
  {\relax}

\bibitem[Schwierz({2010})]{Schwierz_NNano_2010}
Schwierz,~F. {Graphene transistors}. \emph{{Nature Nanotech.}} \textbf{{2010}},
  \emph{{5}}, {487--496}\relax
\mciteBstWouldAddEndPuncttrue
\mciteSetBstMidEndSepPunct{\mcitedefaultmidpunct}
{\mcitedefaultendpunct}{\mcitedefaultseppunct}\relax
\EndOfBibitem
\bibitem[Chen et~al.({2008})Chen, Jang, Xiao, Ishigami, and
  Fuhrer]{Chen_NNano_2008}
Chen,~J.-H.; Jang,~C.; Xiao,~S.; Ishigami,~M.; Fuhrer,~M.~S. {Intrinsic and
  extrinsic performance limits of graphene devices on SiO2}. \emph{{Nature
  Nanotech.}} \textbf{{2008}}, \emph{{3}}, {206--209}\relax
\mciteBstWouldAddEndPuncttrue
\mciteSetBstMidEndSepPunct{\mcitedefaultmidpunct}
{\mcitedefaultendpunct}{\mcitedefaultseppunct}\relax
\EndOfBibitem
\bibitem[Ponomarenko et~al.({2009})Ponomarenko, Yang, Mohiuddin, Katsnelson,
  Novoselov, Morozov, Zhukov, Schedin, Hill, and Geim]{Ponomarenko_PRL_2009}
Ponomarenko,~L.~A. et~al.  {Effect of a High-kappa Environment on Charge
  Carrier Mobility in Graphene}. \emph{{Phys. Rev. Lett.}} \textbf{{2009}},
  \emph{{102}}, {206603}\relax
\mciteBstWouldAddEndPuncttrue
\mciteSetBstMidEndSepPunct{\mcitedefaultmidpunct}
{\mcitedefaultendpunct}{\mcitedefaultseppunct}\relax
\EndOfBibitem
\bibitem[Dean et~al.({2010})Dean, Young, Meric, Lee, Wang, Sorgenfrei,
  Watanabe, Taniguchi, Kim, Shepard, and Hone]{Dean_NNano_2010}
Dean,~C.~R. et~al.  {Boron nitride substrates for high-quality graphene
  electronics}. \emph{{Nature Nanotech.}} \textbf{{2010}}, \emph{{5}},
  {722--726}\relax
\mciteBstWouldAddEndPuncttrue
\mciteSetBstMidEndSepPunct{\mcitedefaultmidpunct}
{\mcitedefaultendpunct}{\mcitedefaultseppunct}\relax
\EndOfBibitem
\bibitem[Xue et~al.({2011})Xue, Sanchez-Yamagishi, Bulmash, Jacquod, Deshpande,
  Watanabe, Taniguchi, Jarillo-Herrero, and LeRoy]{Xue_NMat_2011}
Xue,~J. et~al.  {Scanning tunnelling microscopy and spectroscopy of ultra-flat
  graphene on hexagonal boron nitride}. \emph{{Nature Mater.}} \textbf{{2011}},
  \emph{{10}}, {282--285}\relax
\mciteBstWouldAddEndPuncttrue
\mciteSetBstMidEndSepPunct{\mcitedefaultmidpunct}
{\mcitedefaultendpunct}{\mcitedefaultseppunct}\relax
\EndOfBibitem
\bibitem[Han et~al.({2007})Han, Oezyilmaz, Zhang, and Kim]{Han_RPL_2007}
Han,~M.~Y.; Oezyilmaz,~B.; Zhang,~Y.; Kim,~P. {Energy band-gap engineering of
  graphene nanoribbons}. \emph{{Phys. Rev. Lett.}} \textbf{{2007}},
  \emph{{98}}, {206805}\relax
\mciteBstWouldAddEndPuncttrue
\mciteSetBstMidEndSepPunct{\mcitedefaultmidpunct}
{\mcitedefaultendpunct}{\mcitedefaultseppunct}\relax
\EndOfBibitem
\bibitem[Son et~al.({2006})Son, Cohen, and Louie]{Son_PRL_2006}
Son,~Y.-W.; Cohen,~M.~L.; Louie,~S.~G. {Energy gaps in graphene nanoribbons}.
  \emph{{Phys. Rev. Lett.}} \textbf{{2006}}, \emph{{97}}, {216803}\relax
\mciteBstWouldAddEndPuncttrue
\mciteSetBstMidEndSepPunct{\mcitedefaultmidpunct}
{\mcitedefaultendpunct}{\mcitedefaultseppunct}\relax
\EndOfBibitem
\bibitem[Li et~al.({2011})Li, Qin, Li, Yu, Liu, Luo, Gao, and
  Lu]{Li_ACSNano_2011}
Li,~L. et~al.  {Functionalized Graphene for High-Performance Two-Dimensional
  Spintronics Devices}. \emph{{ACS Nano}} \textbf{{2011}}, \emph{{5}},
  {2601--2610}\relax
\mciteBstWouldAddEndPuncttrue
\mciteSetBstMidEndSepPunct{\mcitedefaultmidpunct}
{\mcitedefaultendpunct}{\mcitedefaultseppunct}\relax
\EndOfBibitem
\bibitem[Abanin et~al.({2010})Abanin, Shytov, and Levitov]{Abanin_PRL_2010}
Abanin,~D.~A.; Shytov,~A.~V.; Levitov,~L.~S. {Peierls-Type Instability and
  Tunable Band Gap in Functionalized Graphene}. \emph{{Phys. Rev. Lett.}}
  \textbf{{2010}}, \emph{{105}}, {086802}\relax
\mciteBstWouldAddEndPuncttrue
\mciteSetBstMidEndSepPunct{\mcitedefaultmidpunct}
{\mcitedefaultendpunct}{\mcitedefaultseppunct}\relax
\EndOfBibitem
\bibitem[Balog et~al.({2010})Balog, Jorgensen, Nilsson, Andersen, Rienks,
  Bianchi, Fanetti, Laegsgaard, Baraldi, Lizzit, Sljivancanin, Besenbacher,
  Hammer, Pedersen, Hofmann, and Hornekaer]{Balog_NMat_2010}
Balog,~R. et~al.  {Bandgap opening in graphene induced by patterned hydrogen
  adsorption}. \emph{{Nature Mater.}} \textbf{{2010}}, \emph{{9}},
  {315--319}\relax
\mciteBstWouldAddEndPuncttrue
\mciteSetBstMidEndSepPunct{\mcitedefaultmidpunct}
{\mcitedefaultendpunct}{\mcitedefaultseppunct}\relax
\EndOfBibitem
\bibitem[Haberer et~al.({2010})Haberer, Vyalikh, Taioli, Dora, Farjam, Fink,
  Marchenko, Pichler, Ziegler, Simonucci, Dresselhaus, Knupfer, Buechner, and
  Grueneis]{Haberer_NanoLetters_2010}
Haberer,~D. et~al.  {Tunable Band Gap in Hydrogenated Quasi-Free-standing
  Graphene}. \emph{{Nano Lett.}} \textbf{{2010}}, \emph{{10}},
  {3360--3366}\relax
\mciteBstWouldAddEndPuncttrue
\mciteSetBstMidEndSepPunct{\mcitedefaultmidpunct}
{\mcitedefaultendpunct}{\mcitedefaultseppunct}\relax
\EndOfBibitem
\bibitem[Elias et~al.({2009})Elias, Nair, Mohiuddin, Morozov, Blake, Halsall,
  Ferrari, Boukhvalov, Katsnelson, Geim, and Novoselov]{Elias_Science_2009}
Elias,~D.~C. et~al.  {Control of Graphene's Properties by Reversible
  Hydrogenation: Evidence for Graphane}. \emph{{Science}} \textbf{{2009}},
  \emph{{323}}, {610--613}\relax
\mciteBstWouldAddEndPuncttrue
\mciteSetBstMidEndSepPunct{\mcitedefaultmidpunct}
{\mcitedefaultendpunct}{\mcitedefaultseppunct}\relax
\EndOfBibitem
\bibitem[Zhou et~al.({2009})Zhou, Wang, Sun, Chen, Kawazoe, and
  Jena]{Zhou_NanoLetters_2009}
Zhou,~J. et~al.  {Ferromagnetism in Semihydrogenated Graphene Sheet}.
  \emph{{Nano Lett.}} \textbf{{2009}}, \emph{{9}}, {3867--3870}\relax
\mciteBstWouldAddEndPuncttrue
\mciteSetBstMidEndSepPunct{\mcitedefaultmidpunct}
{\mcitedefaultendpunct}{\mcitedefaultseppunct}\relax
\EndOfBibitem
\bibitem[Fiori et~al.({2010})Fiori, Lebegue, Betti, Michetti, Klintenberg,
  Eriksson, and Iannaccone]{Fiori_PRB_2010}
Fiori,~G. et~al.  {Simulation of hydrogenated graphene field-effect transistors
  through a multiscale approach}. \emph{{Phys. Rev. B}} \textbf{{2010}},
  \emph{{82}}, {153404}\relax
\mciteBstWouldAddEndPuncttrue
\mciteSetBstMidEndSepPunct{\mcitedefaultmidpunct}
{\mcitedefaultendpunct}{\mcitedefaultseppunct}\relax
\EndOfBibitem
\bibitem[Giovannetti et~al.({2007})Giovannetti, Khomyakov, Brocks, Kelly, and
  van~den Brink]{Giovannetti_PRB_2007}
Giovannetti,~G.; Khomyakov,~P.~A.; Brocks,~G.; Kelly,~P.~J.; van~den Brink,~J.
  {Substrate-induced band gap in graphene on hexagonal boron nitride: Ab initio
  density functional calculations}. \emph{{Phys. Rev. B}} \textbf{{2007}},
  \emph{{76}}, {073103}\relax
\mciteBstWouldAddEndPuncttrue
\mciteSetBstMidEndSepPunct{\mcitedefaultmidpunct}
{\mcitedefaultendpunct}{\mcitedefaultseppunct}\relax
\EndOfBibitem
\bibitem[Subrahmanyam et~al.({2011})Subrahmanyam, Kumar, Maitra, Govindaraj,
  Hembram, Waghmare, and Rao]{Subrahmanyam_PNAS_2011}
Subrahmanyam,~K.~S. et~al.  {Chemical storage of hydrogen in few-layer
  graphene}. \emph{{Proc. Natl. Acad. Sci. USA}} \textbf{{2011}}, \emph{{108}},
  {2674--2677}\relax
\mciteBstWouldAddEndPuncttrue
\mciteSetBstMidEndSepPunct{\mcitedefaultmidpunct}
{\mcitedefaultendpunct}{\mcitedefaultseppunct}\relax
\EndOfBibitem
\bibitem[Neaton et~al.({2006})Neaton, Hybertsen, and Louie]{Neaton_PRL_2006}
Neaton,~J.~B.; Hybertsen,~M.~S.; Louie,~S.~G. {Renormalization of molecular
  electronic levels at metal-molecule interfaces}. \emph{{Phys. Rev. Lett.}}
  \textbf{{2006}}, \emph{{97}}, {216405}\relax
\mciteBstWouldAddEndPuncttrue
\mciteSetBstMidEndSepPunct{\mcitedefaultmidpunct}
{\mcitedefaultendpunct}{\mcitedefaultseppunct}\relax
\EndOfBibitem
\bibitem[Thygesen and Rubio({2009})Thygesen, and Rubio]{Thygesen_PRL_2009}
Thygesen,~K.~S.; Rubio,~A. {Renormalization of Molecular Quasiparticle Levels
  at Metal-Molecule Interfaces: Trends across Binding Regimes}. \emph{{Phys.
  Rev. Lett.}} \textbf{{2009}}, \emph{{102}}, {046802}\relax
\mciteBstWouldAddEndPuncttrue
\mciteSetBstMidEndSepPunct{\mcitedefaultmidpunct}
{\mcitedefaultendpunct}{\mcitedefaultseppunct}\relax
\EndOfBibitem
\bibitem[Freysoldt et~al.({2009})Freysoldt, Rinke, and
  Scheffler]{Freysoldt_PRL_2009}
Freysoldt,~C.; Rinke,~P.; Scheffler,~M. {Controlling Polarization at Insulating
  Surfaces: Quasiparticle Calculations for Molecules Adsorbed on Insulator
  Films}. \emph{{Phys. Rev. Lett.}} \textbf{{2009}}, \emph{{103}},
  {056803}\relax
\mciteBstWouldAddEndPuncttrue
\mciteSetBstMidEndSepPunct{\mcitedefaultmidpunct}
{\mcitedefaultendpunct}{\mcitedefaultseppunct}\relax
\EndOfBibitem
\bibitem[Li et~al.({2009})Li, Lu, and Galli]{Li_JChemTheoComp_2009}
Li,~Y.; Lu,~D.; Galli,~G. {Calculation of Quasi-Particle Energies of Aromatic
  Self-Assembled Monolayers on Au(111)}. \emph{{J. Chem. Theory Comput.}}
  \textbf{{2009}}, \emph{{5}}, {881--886}\relax
\mciteBstWouldAddEndPuncttrue
\mciteSetBstMidEndSepPunct{\mcitedefaultmidpunct}
{\mcitedefaultendpunct}{\mcitedefaultseppunct}\relax
\EndOfBibitem
\bibitem[Gonze et~al.({2009})Gonze, Amadon, Anglade, Beuken, Bottin, Boulanger,
  Bruneval, Caliste, Caracas, Cote, Deutsch, Genovese, Ghosez, Giantomassi,
  Goedecker, Hamann, Hermet, Jollet, Jomard, Leroux, Mancini, Mazevet,
  Oliveira, Onida, Pouillon, Rangel, Rignanese, Sangalli, Shaltaf, Torrent,
  Verstraete, Zerah, and Zwanziger]{Gonze_CompPhysComm_2009}
Gonze,~X. et~al.  {ABINIT: First-principles approach to material and nanosystem
  properties}. \emph{{Computer Physics Communications}} \textbf{{2009}},
  \emph{{180}}, {2582--2615}\relax
\mciteBstWouldAddEndPuncttrue
\mciteSetBstMidEndSepPunct{\mcitedefaultmidpunct}
{\mcitedefaultendpunct}{\mcitedefaultseppunct}\relax
\EndOfBibitem
\bibitem[Troullier and Martins({1991})Troullier, and
  Martins]{Troullier_PRB_1991}
Troullier,~N.; Martins,~J.~L. {Efficient pseudopotentials for plane-wave
  calculations}. \emph{{Phys. Rev. B}} \textbf{{1991}}, \emph{{43}},
  {1993--2006}\relax
\mciteBstWouldAddEndPuncttrue
\mciteSetBstMidEndSepPunct{\mcitedefaultmidpunct}
{\mcitedefaultendpunct}{\mcitedefaultseppunct}\relax
\EndOfBibitem
\bibitem[Goedecker et~al.({1996})Goedecker, Teter, and
  Hutter]{Goedecker_PRB_1996}
Goedecker,~S.; Teter,~M.; Hutter,~J. {Separable dual-space Gaussian
  pseudopotentials}. \emph{{Phys. Rev. B}} \textbf{{1996}}, \emph{{54}},
  {1703--1710}\relax
\mciteBstWouldAddEndPuncttrue
\mciteSetBstMidEndSepPunct{\mcitedefaultmidpunct}
{\mcitedefaultendpunct}{\mcitedefaultseppunct}\relax
\EndOfBibitem
\bibitem[Hybertsen and Louie({1986})Hybertsen, and Louie]{Hybertsen_PRB_1986}
Hybertsen,~M.~S.; Louie,~S. {Electron correlation in semiconductors and
  insulators: Band gaps and quasiparticle energies}. \emph{{Phys. Rev. B}}
  \textbf{{1986}}, \emph{{34}}, {5390--5413}\relax
\mciteBstWouldAddEndPuncttrue
\mciteSetBstMidEndSepPunct{\mcitedefaultmidpunct}
{\mcitedefaultendpunct}{\mcitedefaultseppunct}\relax
\EndOfBibitem
\bibitem[Kresse and Furthm\"uller(1996)Kresse, and
  Furthm\"uller]{kresse1996_VASP1}
Kresse,~G.; Furthm\"uller,~J. Efficiency of ab-initio total energy calculations
  for metals and semiconductors using a plane-wave basis set.
  \emph{Computational Materials Science} \textbf{1996}, \emph{6}, 15--50\relax
\mciteBstWouldAddEndPuncttrue
\mciteSetBstMidEndSepPunct{\mcitedefaultmidpunct}
{\mcitedefaultendpunct}{\mcitedefaultseppunct}\relax
\EndOfBibitem
\bibitem[Bucko et~al.({2010})Bucko, Hafner, Lebegue, and
  Angyan]{Bucko_JPCA_2010}
Bucko,~T.; Hafner,~J.; Lebegue,~S.; Angyan,~J.~G. {Improved Description of the
  Structure of Molecular and Layered Crystals: Ab Initio DFT Calculations with
  van der Waals Corrections}. \emph{{Journal of Physical Chemistry A}}
  \textbf{{2010}}, \emph{{114}}, {11814--11824}\relax
\mciteBstWouldAddEndPuncttrue
\mciteSetBstMidEndSepPunct{\mcitedefaultmidpunct}
{\mcitedefaultendpunct}{\mcitedefaultseppunct}\relax
\EndOfBibitem
\bibitem[Kresse and Joubert({1999})Kresse, and
  Joubert]{Kresse_PRB_1999_VASP_PAW}
Kresse,~G.; Joubert,~D. {From ultrasoft pseudopotentials to the projector
  augmented-wave method}. \emph{{Phys. Rev. B}} \textbf{{1999}}, \emph{{59}},
  {1758--1775}\relax
\mciteBstWouldAddEndPuncttrue
\mciteSetBstMidEndSepPunct{\mcitedefaultmidpunct}
{\mcitedefaultendpunct}{\mcitedefaultseppunct}\relax
\EndOfBibitem
\bibitem[Perdew et~al.({1996})Perdew, Burke, and
  Ernzerhof]{Perdew_PRL_1996_PBE}
Perdew,~J.; Burke,~K.; Ernzerhof,~M. {Generalized gradient approximation made
  simple}. \emph{{Phys. Rev. Lett.}} \textbf{{1996}}, \emph{{77}},
  {3865--3868}\relax
\mciteBstWouldAddEndPuncttrue
\mciteSetBstMidEndSepPunct{\mcitedefaultmidpunct}
{\mcitedefaultendpunct}{\mcitedefaultseppunct}\relax
\EndOfBibitem
\bibitem[Grimme({2006})]{Grimme_JCC_2006}
Grimme,~S. {Semiempirical GGA-type density functional constructed with a
  long-range dispersion correction}. \emph{{Journal of Computational
  Chemistry}} \textbf{{2006}}, \emph{{27}}, {1787--1799}\relax
\mciteBstWouldAddEndPuncttrue
\mciteSetBstMidEndSepPunct{\mcitedefaultmidpunct}
{\mcitedefaultendpunct}{\mcitedefaultseppunct}\relax
\EndOfBibitem
\bibitem[Shallcross et~al.({2008})Shallcross, Sharma, and
  Pankratov]{PRL_Shallcross_2008}
Shallcross,~S.; Sharma,~S.; Pankratov,~O.~A. {Quantum interference at the twist
  boundary in graphene}. \emph{{Phys. Rev. Lett.}} \textbf{{2008}},
  \emph{{101}}, {056803}\relax
\mciteBstWouldAddEndPuncttrue
\mciteSetBstMidEndSepPunct{\mcitedefaultmidpunct}
{\mcitedefaultendpunct}{\mcitedefaultseppunct}\relax
\EndOfBibitem
\bibitem[Sofo et~al.({2007})Sofo, Chaudhari, and Barber]{Sofo_PRB_2007}
Sofo,~J.~O.; Chaudhari,~A.~S.; Barber,~G.~D. {Graphane: A two-dimensional
  hydrocarbon}. \emph{{Phys. Rev. B}} \textbf{{2007}}, \emph{{75}},
  {153401}\relax
\mciteBstWouldAddEndPuncttrue
\mciteSetBstMidEndSepPunct{\mcitedefaultmidpunct}
{\mcitedefaultendpunct}{\mcitedefaultseppunct}\relax
\EndOfBibitem
\bibitem[Zhou et~al.({2009})Zhou, Wu, Zhou, and Sun]{Zhou_APL_2009}
Zhou,~J.; Wu,~M.~M.; Zhou,~X.; Sun,~Q. {Tuning electronic and magnetic
  properties of graphene by surface modification}. \emph{{Appl. Phys. Lett.}}
  \textbf{{2009}}, \emph{{95}}, {103108}\relax
\mciteBstWouldAddEndPuncttrue
\mciteSetBstMidEndSepPunct{\mcitedefaultmidpunct}
{\mcitedefaultendpunct}{\mcitedefaultseppunct}\relax
\EndOfBibitem
\bibitem[Jiang et~al.()Jiang, Kharche, and Nayak]{Jiang_to_be_published}
Jiang,~X.; Kharche,~N.; Nayak,~S.~K. (to be published). \relax
\mciteBstWouldAddEndPunctfalse
\mciteSetBstMidEndSepPunct{\mcitedefaultmidpunct}
{}{\mcitedefaultseppunct}\relax
\EndOfBibitem
\bibitem[Pan et~al.({1998})Pan, Wilks, Dunstan, Pritchard, Williams, Cammack,
  and Clark]{Pan_APL_1998}
Pan,~M. et~al.  {Modification of band offsets by a ZnSe intralayer at the
  Si/Ge(111) interface}. \emph{{Appl. Phys. Lett.}} \textbf{{1998}},
  \emph{{72}}, {2707--2709}\relax
\mciteBstWouldAddEndPuncttrue
\mciteSetBstMidEndSepPunct{\mcitedefaultmidpunct}
{\mcitedefaultendpunct}{\mcitedefaultseppunct}\relax
\EndOfBibitem
\bibitem[Robertson({2000})]{Robertson_JVCTB_2000}
Robertson,~J. {Band offsets of wide-band-gap oxides and implications for future
  electronic devices}. \emph{{J. Vac. Sci. Technol. B}} \textbf{{2000}},
  \emph{{18}}, {1785--1791}\relax
\mciteBstWouldAddEndPuncttrue
\mciteSetBstMidEndSepPunct{\mcitedefaultmidpunct}
{\mcitedefaultendpunct}{\mcitedefaultseppunct}\relax
\EndOfBibitem
\end{mcitethebibliography}

% Copy contents of file GNR_stacking_bibliography.bbl before submitting for publication.
\providecommand*\mcitethebibliography{\thebibliography}
\csname @ifundefined\endcsname{endmcitethebibliography}
  {\let\endmcitethebibliography\endthebibliography}{}

%----------------------------- Bibliography --------------------------------

\end{document}